\documentclass[fleqn,usenatbib]{mnras}

\usepackage{newtxtext,newtxmath}

\usepackage[T1]{fontenc}

\DeclareRobustCommand{\VAN}[3]{#2}
\let\VANthebibliography\thebibliography
\def\thebibliography{\DeclareRobustCommand{\VAN}[3]{##3}\VANthebibliography}

\usepackage{graphicx}	
\usepackage{caption}
\usepackage{subcaption}
\usepackage{amsmath}	
\usepackage{bbold}
\usepackage{multicol}        
\usepackage{bm}		
\usepackage{pdflscape}	
\usepackage{multirow}
\usepackage{booktabs}
\usepackage{graphicx}
\usepackage{array}
\usepackage{float}
\usepackage{xspace} 
\usepackage{subfiles} 
\usepackage{makecell}

\newcommand{\inv}{^{\raisebox{.2ex}{$\scriptscriptstyle-1$}}}
\newcommand{\mat}[1]{\mathbf{#1}}

\newcommand{\Planck}{\emph{Planck}\xspace}

\newcommand{\madam}{\texttt{Madam}\xspace}

\newcommand{\SRtwo}{\texttt{SRoll2}\xspace}

\newcommand{\glass}{\textsc{glass}\xspace}
\newcommand{\momento}{\texttt{momento}\xspace}

\newcommand{\NPIPE}{\texttt{NPIPE}\xspace}
\newcommand{\pixlike}{\texttt{pixLike}\xspace}

\def\ltsima{$\; \buildrel < \over \sim \;$\xspace}
\def\gtsima{$\; \buildrel > \over \sim \;$\xspace}
\def\simlt{\lower.5ex\hbox{\ltsima}}
\def\simgt{\lower.5ex\hbox{\gtsima}}

\newcommand{\kpivot}{0.05}
\newcommand{\rNRPL}{$r_{\kpivot}<0.274$\xspace}
\newcommand{\rNR}{$r_{\kpivot}<0.408$\xspace}

\newcommand{\CL}{(95\% C.L.)\xspace}

\providecommand{\sorthelp}[1]{}
\usepackage[dvipsnames]{xcolor}

\newcolumntype{N}{>{\centering\arraybackslash}m{.5in}}
\newcolumntype{G}{>{\centering\arraybackslash}m{2in}}

\defcitealias{2016A&A...596A.107P}{PSRoll1}
\defcitealias{Aghanim:2018eyx}{PCP18}
\defcitealias{2020A&A...641A...5P}{PLP19}
\defcitealias{2019A&A...629A..38D}{PSRoll2}
\defcitealias{EG19}{EG}
\defcitealias{Belsunce21}{B21}
\defcitealias{Belsunce22}{B22}
\defcitealias{BICEP:2021}{BK18}
\defcitealias{HL:2008}{HL08}

\title[Low-$\ell$ $B$-mode constraints from Planck]{$B$-mode constraints from \Planck low multipole polarisation  data}

\author[de Belsunce, Gratton \& Efstathiou]{Roger de Belsunce,\thanks{E-mail: \href{mailto:rmvd2@cam.ac.uk}{rmvd2@cam.ac.uk}} Steven Gratton, and George Efstathiou
\\Kavli Institute for Cosmology \& Institute of Astronomy, University of Cambridge, Madingley Road, Cambridge CB3 OHA, United Kingdom}

\date{Accepted XXX. Received YYY; in original form ZZZ}

\pubyear{2022}

\begin{document}
\label{firstpage}
\pagerange{\pageref{firstpage}--\pageref{lastpage}}
\maketitle

\begin{abstract}
We present constraints on primordial $B$ modes from large angular scale cosmic microwave background polarisation anisotropies measured with the \Planck satellite. To remove Galactic polarised foregrounds, we use a Bayesian parametric component separation method, modelling synchrotron radiation as a power law and thermal dust emission as a modified blackbody. This method propagates uncertainties from the foreground cleaning into the noise covariance matrices of the maps. We construct two likelihoods: (i) a semi-analytical cross-spectrum-based likelihood-approximation scheme (\momento) and (ii) an exact polarisation-only pixel-based likelihood (\pixlike). Since \momento is based on cross-spectra it is statistically less powerful than \pixlike, but is less sensitive to systematic errors correlated across frequencies. Both likelihoods give a tensor-to-scalar ratio, $r$, that is consistent with zero from low multipole ($2\leq \ell < 30$) \Planck polarisation data. From full-mission maps we obtain \rNRPL, at $95\%$ confidence, at a pivot scale of $k_{\ast} \equiv \kpivot\ \mathrm{Mpc}^{-1}$, using \pixlike. \momento gives a qualitatively similar but weaker $95\%$ confidence limit of \rNR.
\end{abstract}




\section{Introduction} \label{sec:intro}
Measurements of the polarised cosmic microwave background (CMB) on large angular scales provide a unique opportunity to constrain the physics of the early Universe. In particular, a detection of primordial $B$-modes, parameterised by the tensor-to-scalar ratio $r$,  would provide strong evidence that the Universe underwent an inflationary phase and would shed light on the dynamics involved as follows.  

Scalar perturbations generated in the early Universe yield a divergence-like `$E$-mode' polarisation pattern in the CMB, whereas tensor perturbations give an additional distinctive curl-like `$B$-mode' pattern  \citep{Seljak:1997, Kamionkowski:1997}, of approximately equal amplitude to their $E$-mode contribution. A definitive  detection of primordial  $B$-modes would provide strong constraints on  models of inflation and would fix the energy scale of inflation through the relation \citep{Lyth:1985}
\begin{equation}
    V^{1/4} \approx 1.04 \times 10^{16}\ \mathrm{GeV} \left( \frac{r}{0.01}\right)^{1/4} \ ,
\end{equation}
where $V$ is the potential energy of the inflaton and $r$ the ratio of the amplitudes of the tensor and scalar power spectra at an arbitrary pivot scale $k_{\ast} \equiv \kpivot\ \mathrm{Mpc}^{-1}$. 

One of the main science goals of next-generation CMB experiments from space \citep[LiteBIRD;][]{2020JLTP..199.1107S} and from the ground, such as the Atacama Cosmology Telescope \citep[ACTPol,][]{Thornton:2016}, South Pole Telescope \citep[SPTpol,][]{Austermann2012}, CMB-S4 \citep{CMBS4:2020} and the Simons Observatory \citep[SO,][]{SO_2019} is  to achieve a statistically significant detection of primordial $B$-modes down to $\sigma_r \sim 10^{-3}$. Galactic polarised foregrounds, noise as well as instrumental systematics, however, pose a real challenge for this ambitious science goal. 

Currently, the tightest constraint on the tensor-to-scalar ratio  comes from  CMB polarization measurements by the BICEP/{\it Keck} experiment  on scales  of the \emph{recombination peak} \citep[][hereafter \citetalias{BICEP:2021}]{BICEP:2021}. \citetalias{BICEP:2021} measures $r$ at intermediate multipoles ($20 < \ell < 330$) using roughly 400 deg$^2$ of the sky. They place tight constraints on $r_{\kpivot}< 0.036$ \CL, when fixing the $\Lambda$CDM cosmological model to \Planck 2018 best-fit values \citep[][hereafter \citetalias{Aghanim:2018eyx}]{Aghanim:2018eyx} and using a pivot scale\footnote{To be consistent with \citetalias{BICEP:2021}, we use a pivot scale $k_{\ast}=\kpivot\ \mathrm{Mpc}^{-1}$ in this paper also. Note that \citetalias{Aghanim:2018eyx} uses $k_{\ast} = 0.002\ \mathrm{Mpc}^{-1}$ as pivot scale. Assuming a scaling relation of $(0.05/0.002)^{-r/8}$ and adopting the single-field-inflation consistency relation ($n_t=-r/8$), we estimate the deviations in $r$ for different $k_{\ast}$ to be $\sim 5-10 \%$ for $r\sim 0.1-0.3$.} of $k_{\ast}=\kpivot\ \mathrm{Mpc}^{-1}$. The recent analysis of the first flight of the SPIDER balloon-borne experiment gives a weaker, but independent, 95\% confidence limit of $r_{0.05} < 0.19$ from the recombination peak \citep{SPIDER:2022}. 

Since the \Planck satellite observed the full sky, the \Planck\ polarization  maps contain information in the low multipole regime ($2\leq \ell < 30$).  Thus, with \Planck one has potential access to the \emph{reionization bump} feature caused by tensor perturbations\footnote{See \cite{2009JCAP...06..011E} for a pre-launch analysis.}. It is important to have a reliable analysis 
of the \Planck\ $B$-mode constraints, since there is no other experiment at present that can access these large angular scales (which are unconstrained by  \citetalias{BICEP:2021}). $B$-mode observations over a  wide range of angular scales can provide tests of inflationary dynamics, for example, whether there was
an early period of fast roll inflation \citep[see e.g.][]{Contaldi:2014}. The goal of the present paper is to establish an accurate upper limit on $r$ from low-$\ell$ \Planck polarisation data.

Recently, tight constraints on $r$ using \Planck low-$\ell$ maps and simulations from the  \NPIPE processing pipeline \citep{NPIPE:2020} have been presented in \citet{Tristram:2021a,Tristram:2021b}. Using the reionization bump $BB$ spectrum in the multipole range $2\leq \ell \leq 35$, and allowing negative values of $r_{0.05}$, they found the constraint 
\begin{align}
r_{0.05}^{BB}   &= -0.014^{+0.108}_{-0.111}\qquad \text{\citep{Tristram:2021b}}   \label{eq:Tristram_r}
\end{align}
at the $95\%$ confidence level (C.L.).  Extending the multipole range to  $2\leq \ell \leq 150$, thus including $BB$ data from the recombination bump,  they found $r_{0.05}<0.158$ \CL. Combining temperature and polarisation data and adding a high-$\ell$ CMB temperature likelihood gave a surprisingly tight upper limit of $r_{0.05}<0.056$ \CL  \citep{Tristram:2021b}. Combining the \NPIPE and \citetalias{BICEP:2021} likelihoods \cite{Tristram:2021b} find  $r_{0.05}<0.032$ \CL, i.e.~the addition of \Planck\ $BB$ produces a negligible improvement to constraints from   \citetalias{BICEP:2021}.  The  \citet{Tristram:2021b} results disagree strongly with the previous analysis of \Planck polarization maps in \citet{2019arXiv190712875P}, which found  a relatively weak constraint of $r_{0.002} < 0.41$ \CL from \Planck\ $BB$ measurements at low multipoles. This is much weaker than the \Planck\ constraint on $r$ from $TT$ measurements alone of $r_{0.002} < 0.102$ \citepalias[][]{Aghanim:2018eyx}.

At low multipoles, the \cite{Tristram:2021b} analysis uses a heuristic low-multipole $EE$, $BB$, $EB$ likelihood 
motivated by the Hammimeche-Lewis ansatz  \citep[][hereafter \citetalias{HL:2008}]{HL:2008} with covariance matrices computed from 
400 end-to-end numerical simulations.
A recent analysis presented in \citet{Beck:2022}, however, shows that inaccuracies in simulation-based noise covariance matrices (NCMs) 
can bias results towards low values of $r$. Further, \citet{Beck:2022} repeated the \Planck\ analysis  of \citet{Tristram:2021b} by `conditioning' the numerical
covariance matrices (i.e.\ setting off-diagonal terms to zero) finding significantly weaker constraints of $r_{0.05}<0.13$ \CL when combined with a high-$\ell$ likelihood. The weaker constraint comes 
primarily from the impact of conditioning on the \cite{Tristram:2021b} low multipole likelihood.  \citet{Campeti:2022} found similar results to \citet{Beck:2022} from a profile likelihood analysis.

In this paper, our aim is to reassess the \Planck\ constraint on $r$ using  \textit{only the low `reionization' range of multipoles}  ($2 \leq \ell < 30$). It is important to establish a reliable limit from this multipole range, since it is not covered by  \citetalias{BICEP:2021} 
(which has much higher signal-to-noise than \Planck\ over the multipole range of the recombination feature and so the  \Planck\ $B$-mode constraints over this  multipole range offer little more than a crude consistency
test  of \citetalias{BICEP:2021}).

Determining $r$ from large angular scale CMB data is challenging. Galactic polarised foregrounds need to be subtracted to high accuracy and
instrument noise and systematics reduced to low levels to be able to set
interesting limits on primordial $B$-modes. In this paper we use NCMs that were introduced in \citet[][hereafter \citetalias{Belsunce21}]{Belsunce21}, capturing aspects of systematic effects from end-to-end simulations. We employ the Bayesian parametric component separation method from \citet[][hereafter \citetalias{Belsunce22}]{Belsunce22}, instead of subtracting  single coefficient foreground  templates.  This foreground removal scheme propagates uncertainties from  cleaning through to effective NCMs for the CMB, which are then used in the likelihoods. We constrain the tensor-to-scalar ratio using two approaches: (i) a semi-analytical cross-spectrum-based likelihood-approximation scheme \momento\ using all available polarisation cross-spectra between the half-mission \Planck maps, namely $EE,\ EB,\ BE$ and $BB$, and (ii) a polarisation-only pixel-based likelihood (\pixlike) using full-mission polarised \Planck maps. Each of these likelihood schemes has a sound theoretical basis.

We assume adiabatic perturbations, with primordial power spectra for the scalar ($s$) and tensor ($t$) modes given by
\begin{subequations}
\begin{align} \label{eq:theory_pk}
\mathcal{P}_s(k) &= A_s \left(\frac{k}{k_{\ast}}\right)^{n_s - 1}\ , \\
\mathcal{P}_t(k) &= A_t \left(\frac{k}{k_{\ast}}\right)^{n_t}\ , 
\end{align}
\end{subequations}
with constant spectral indices (i.e.\ no `running', $\mathrm{d}n_{s,t}/\mathrm{d}\ln k = 0$). The tensor-to-scalar ratio $r$ is defined as the ratio of the perturbation amplitudes, $r\equiv A_t/A_s$, at some chosen pivot scale $k_{\ast} \equiv \kpivot\ \mathrm{Mpc}^{-1}$. As in \citetalias{Aghanim:2018eyx}, we fix the tensor spectral index in  terms of the tensor-to-scalar-ratio via the single-field-inflation consistency relation $n_t = -r/8$.

This paper is organised as follows. Section\ \ref{sec:data} discusses the \Planck data that we employ along with the required pixel-pixel noise covariance matrices. The likelihoods are introduced in Sec.\ \ref{sec:likelihoods}. Tests of the likelihoods on simulations are given in Sec.\ \ref{sec:results_sims}. Results using \Planck polarisation data are presented in Sec.\ \ref{sec:results} and  our conclusions are 
summarised in  Sec.\ \ref{sec:conclusion}. Two appendices address aspects of the noise modelling.

\section{Data} \label{sec:data}
We use the \Planck 2018 Low Frequency Instrument (LFI) maps, as processed with the \madam map-making algorithm and \Planck\ High Frequency Instrument (HFI) data, as processed with the \SRtwo map-making algorithm. The \madam and \SRtwo map-making algorithms are described in detail in \citet{Planck_LFI:2016, LFI_data:2018} and \citet{2019A&A...629A..38D}. Recently, two new sets of \Planck maps have been introduced, \citep[][hereafter \NPIPE] {NPIPE:2020} and \citep[][hereafter \texttt{BeyondPlanck}]{BeyondPlanck:2020}, for LFI and HFI jointly and for LFI alone respectively.
We extend our analysis to \NPIPE data products in Sec.~\ref{sec:NPIPE_results}.\footnote{The LFI maps and simulations (labelled FFP10) are available at \url{https://pla.esac.esa.int}, LFI and HFI \NPIPE data is available at \url{https://portal.nersc.gov/project/cmb/planck2020/}, and HFI \SRtwo products are available at \url{http://sroll20.ias.u-psud.fr}.}

\subsection{Map compression}\label{sec:CMB_data}
We are interested in a signal that affects the polarised CMB anisotropies on large angular scales corresponding to  multipoles $2\leq \ell \leq 30$. Therefore, we degrade the high-resolution \Planck maps with HEALPix parameter $N_{\rm side}=2048$ (corresponding to $5.03 \times 10^7$ pixels) and perform the relevant matrix and map computations at low resolution (lr) with $N_{\rm side}^{\rm lr}=16$ (corresponding to $3072$ pixels). This makes our analysis  computationally tractable. In this paper, we closely follow the analysis methodology  described in \citetalias{Belsunce21} and \citetalias{Belsunce22} and so will describe  the key steps briefly here.

We first degrade the high resolution \Planck $Q$ and $U$ maps. To do this,  we apply an apodised mask\footnote{See Fig.\ 1 in \citetalias{Belsunce21} for illustrations of the masks we use.} with $f_{\rm sky}=0.85$ to suppress the Galactic plane region. (This procedure is required for the contaminated CMB sky maps to avoid smearing high amplitude foreground emission in the Galactic plane to high Galactic latitudes.) The maps are then smoothed using the harmonic-space smoothing operator
\begin{equation} \label{eq:harmonic_space_operator}
w(\ell) = \left\{ \begin{array}{ll}
 1 &\mathrm{for}\ \ell \leq \ell_1\ ,\\
 \frac{1}{2}\left[ 1 + \cos{\pi\frac{\ell - \ell_1}{\ell_2-\ell_1}} \right] &\mathrm{for}\ \ell_1 < \ell \leq \ell_2\ ,\\
 0 &\mathrm{for}\ \ell > \ell_2\ ,
\end{array} \right. \end{equation}
with $\ell_1 = N^{\rm lr}_{\rm side}$ and $\ell_2 = 3N^{\rm lr}_{\rm side}$. We apply the smoothing operator $w(\ell)$ and a $\texttt{HEALPix}$ pixel window function at the map level to remove small-scale power, as discussed in \citetalias{Belsunce21}.

\subsection{Pixel-pixel noise covariance matrices}\label{sec:NCM}
The fidelity of our posteriors on cosmological parameters depends on the noise and systematics modelling entering into the likelihoods. Therefore, we fit a NCM at each frequency to end-to-end simulations, introducing large-scale modes to capture systematics. Such modes are required to describe large-scale polarisation data from HFI, which is affected by analogue-to-digital-converter non-linearities. The fitting scheme has been tested in the context of large-angular scale CMB inference for the optical depth to reionization parameter in \citetalias{Belsunce21}.  

In more detail, we fit parametric models to sample covariance matrices computed from a number $n_s$ of end-to-end simulations of \madam (for LFI) and \SRtwo (for HFI) data, assuming the noise between frequencies to be independent. We omit 100 end-to-end simulations from the fitting, reserving them to test our likelihood approximations.  The sample noise- and systematics-only covariance matrix is constructed as $\mathbf{\Hat{N}} = \sum_i (\mathbf{n}_i - \mathbf{\Bar{n}})(\mathbf{n}_i - \mathbf{\Bar{n}})^{\top} /n_s$, with the overbar denoting average over simulations and $\mathbf{n}_i$ are the simulated sky maps for $Q$ and $U$ Stokes parameters. A smoothed template $\mathbf{\Bar{n}}$ from the simulation maps\footnote{This corresponds to a smoothed mean of the maps, i.e.\ the maximum likelihood solution of the map-making equation, see e.g. \cite{Tegmark:1997}.} is first subtracted when computing this covariance. 

We approach the problem of fitting a model to each $\mathbf{\Hat{N}}$ as a maximum likelihood inference problem. We assume a Gaussian probability distribution for each noise realisation
\begin{align} \label{eq:likelihood_ncm}
    \mathcal{L} \equiv \mathbf{\mathcal{P}}(\mathbf{\hat{N}} \vert \mathcal{\mathbf{M}}) &= \prod_{i=1}^{n_s} \frac{1}{\sqrt{|2\pi \mathbf{M}|}}e^{-\frac{1}{2}(\mathbf{n}_i - \mathbf{\Bar{n}}) \mathbf{M}\inv(\mathbf{n}_i-\mathbf{\Bar{n}})^{\top}}  \ , \end{align}
where  $\mat M$  is the model for the  noise covariance matrix. We take $\mat M$ to consist of two terms 
\begin{equation}
    \mat M  =  \alpha \mat N_0 + \mat{Y\mathbf{\Psi}Y}^\top\ .
\end{equation}
Here the parameter $\alpha$ scales the low resolution (lr) map-making covariance matrix $\mat N_{0}$ and $\mat{Y \mathbf{\Psi}Y}^\top$ models additional large-scale effects. The lr covariance matrices $\mat N_{0}$ were constructed as part of the \Planck\ FFP8 end-to-end simulation effort. They contain structure representing the scanning strategy, detector white noise and `1/f'-type noise and are used as the starting point for our fitting procedure.The additional term $\mat{Y\mathbf{\Psi}Y}^\top$ is an outer product of low-multipole spherical harmonics $\mat{Y}$ weighted with covariance $\mathbf{\Psi}$. We solve analytically for $\mathbf{\Psi}$ as a function of $\alpha$; see Eq. (29) in \citetalias{Belsunce21}. To solve for $\mat{M}$, we minimise the negative log-likelihood from Eq.~\eqref{eq:likelihood_ncm} with respect to $\alpha$ and hence solve for the desired NCM. We refer the reader to \citetalias{Belsunce21} for details. 

\subsection{Galactic polarised foreground cleaning}
Foregrounds pose a significant challenge for the measurement of low amplitude primordial $B$-modes. In the present paper, we use a \emph{Bayesian parametric component separation} method.  We assume that the main Galactic polarised foregrounds, namely synchrotron and thermal dust radiation, have spectral energy distributions that can be
described with a small number of parameters, see \citetalias{Belsunce22} for a detailed analysis in the context of \Planck data.

We assume that the observed sky can be modelled as $\mat d = \mat{Am} + \mat n$ where the measured data vector $\mat d$ consists of stacked sky maps. The sky signal $\mat{Am}$ consists of a scaling matrix $\mat A$ and the model $\mat m$. The instrumentation and measurement noise is $\mat n$. The sky signal encodes the primordial CMB together with our foreground modelling, namely a power law for synchrotron and a modified blackbody emission law for thermal dust. Eq.\ (3) in \citetalias{Belsunce22} summarises our foreground model.

We can disentangle Galactic foregrounds from the primordial CMB and propagate the uncertainties from the cleaning procedure through to the NCMs and the subsequent parameter constraints. Therefore, we estimate the posterior distribution of the sky signal $\mat{Am}$ given the observed data $\mat d$ using a Bayesian framework $\mathcal{P}(\mat{A},\mat{m}\vert \mat{d})$. The procedure develops iteratively, performing a multidimensional Newton-Raphson (NR) minimisation on the spectral index parameters; henceforth we will call this procedure the NR method. 

A key aspect of our approach is the introduction of a spatial correlation length for the spectral indices, penalizing variations below this scale but allowing variations on larger scales across the sky. The induced small-scale rigidity ``regularizes" the recovered component map solutions, reducing any overfitting of noise. The total covariance matrix outputted from the procedure contains correlations both within and between the model components, naturally incorporating information from the pixel-pixel frequency noise correlation matrices mentioned in Sec.\ \ref{sec:NCM} above. 

Our main constraints on $r$ in the present paper will be derived using NR-cleaned CMB maps: from LFI we use the 30 and 70 GHz \madam frequency channels and from HFI we use the 100, 143, 217 and 353 GHz \SRtwo frequency channels. For inference in the map domain, we use full-mission maps. For the likelihood-approximation scheme, we measure the quadratic cross spectra (QCS) between CMB maps made from differing half-mission (HM1/HM2) sets of \Planck\ maps.

\section{Likelihoods} \label{sec:likelihoods}
We now briefly introduce and compare the two likelihoods that we shall use to constrain the tensor-to-scalar ratio from low-$\ell$ \Planck data.  In Sec.\ \ref{sec:Likelihood_GLASS} we present a semi-analytic likelihood approximation scheme for cross-spectrum estimators based on the principle of maximum entropy, called \momento.
In Sec.\ \ref{sec:Likelihood_pixlike} we present an ``exact" polarisation-only pixel-based likelihood, called \pixlike. Effects of regularising noise are discussed in Appendix \ref{app:reg_noise}.

\subsection{Cross-spectrum likelihood-approximation scheme} \label{sec:Likelihood_GLASS}
Our first likelihood is \momento, built according to the principles of the ``General Likelihood Approximate Solution Scheme" \citep[\glass;][]{Gratton:2017}. The main rationale is to only use aspects of the data that one is relatively confident in, here quadratic cross-spectra, when constructing a likelihood.  \glass\ is then useful precisely because it allows one to compute posteriors where exact analytic likelihoods for such statistics are either unknown or too expensive to compute.  \momento\ was extensively tested on simulations and data for the inference of the optical depth to reionization, $\tau$, from low-$\ell$ CMB temperature and polarisation data, as presented in \citetalias{Belsunce21}. \glass\ may be thought of as approximating the sampling distribution of the chosen data as the least presumptive one consistent with a chosen set of the moments of the data, according to the principle of maximum entropy.  The moments may ideally be obtained analytically from the model in question but also could be obtained from forward simulations.  If the models considered are continuous functions of some parameters $a_\alpha$ say, one need not explicitly construct this distribution. Rather, one can actually compute an approximation of its gradient:
\begin{equation}
    -\left(\log\mathcal{L}\right)_{,a_\alpha} = - (X - \langle X^\top \rangle)\langle \langle XX^\top \rangle \rangle \inv \langle X \rangle_{,a_\alpha} \ .\label{eq:dsda}
\end{equation}
This procedure is generally applicable to a multitude of statistics $x^i,\ i \in \mathopen[1,\dots, n\mathclose]$, and $X$ above denotes a vector of these statistics. The $\langle \dots \rangle$ denotes the ensemble average of the enclosed quantity, while  $\langle \langle \dots \rangle \rangle$ denotes the appropriate cumulant or connected moment. One can then integrate along a path in parameter space to obtain the difference in likelihood between any two models.   We refer the reader to \citet{Gratton:2017} and \citetalias{Belsunce21} for a detailed description of the method.

To summarise, \momento\ uses near-optimal quadratic cross-spectra (QCS), to compress the CMB data down to a set of power spectrum elements (here we typically consider multipoles $\ell=2-30$). A fiducial angular power spectrum is needed, $C_\ell^{\mathrm{fid}}$, in order to compute the QCS of the data. Subsequently, at each step in the likelihood the following steps are taken by \momento:
\begin{enumerate}
\item take input set of theory $C_\ell$'s;
\item compute the residual between fiducial and theory spectra $\Delta C_\ell =
  C_\ell-C_\ell^{\mathrm{fid}}$;
 \item choose the path $C_\ell^{\mathrm{fid}}  + \gamma \Delta C_\ell$, with $\gamma \in \left[ 0,1 \right]$;
\item evaluate the gradients of the log-likelihood using Eq.\ \eqref{eq:dsda} from computed moments, as a function of $\gamma$;
\item perform a numerical integration using these gradients to evaluate the change in log-likelihood going from $C_\ell^{\mathrm{fid}}$ to $C_\ell$.
\end{enumerate}
In the present paper, we constrain $r$ with \momento by using all available QCS polarisation spectra: $EE$, $EB$, $BE$ and $BB$ on foreground-cleaned \Planck polarisation maps.

\subsection{Low multipole pixel-based likelihood} \label{sec:Likelihood_pixlike}
Under the assumption that signal and noise are Gaussian and the noise covariance matrix is known accurately, it is possible to write down a pixel-based likelihood for low resolution CMB maps, see e.g.\ \citet{Page:2007}. This likelihood is exact in real space, even for incomplete sky coverage. Although it is theoretically well motivated to assume that  primordial CMB signal is inherently Gaussian, noise properties and systematics may  lead to a  break down of Gaussian assumption. Nevertheless,  since a pixel based likelihood is easy to compute, it provides a  cross-check of the \momento\ scheme.    
Pixel-based likelihoods have two further disadvantages: (i) the NCMs need
to be known accurately; (ii)  the contribution of individual multipoles to the $r$ measurement cannot be isolated or visualized.

The pixel-based polarisation-only likelihood is given by the probability of the data given the theoretical noise and signal model:
\begin{equation} \label{eq:pix_like}
    \mathcal{L}(\mat{d}|\mat{\Omega}) \equiv \mathcal{P}(\mat{d}|\mat{C}) = \frac{1}{\sqrt{|2\pi \mat{C}|}}\exp\left( -\frac{1}{2}(\mat{d}-\overline{\mat{d}})^\top \mat{C}\inv (\mat{d}-\overline{\mat{d}})\right) \ ,
\end{equation}
where $\mat{d}$ is the data vector and $\overline{\mat{d}}$ its smoothed mean (a map-based systematics offset determined from fits to the end-to-end simulations), both containing the polarisation $Q$ and $U$ maps. The model parameters are denoted by $\mat{\Omega}$. The covariance matrix $\mat{C}$ consists of a signal $\mat{S}$ and noise $\mat{N}$ component: $\mat{C} = \mat{S} + \mat{N}$. The signal matrix can be constructed following the procedure outlined in e.g.\ \citet{Tegmark2001}.

Computing the pixel-based likelihood given in Eq.\ \eqref{eq:pix_like} requires calculating the signal covariance matrix as a function of the  theoretical power spectra $C_{\ell}^{\rm th}$ in question. For each \pixlike evaluation at some fiducial set of parameters $\mathcal{L}(\mat{d}|\mat{\Omega})$ the determinant and the inverse of the NCM are computed. However, both operations are of complexity $\mathcal{O}(N_{\rm pix}^3)$ rendering the problem computationally expensive and only feasible at low resolution. 

For our default constraints from the pixel-based method we use the output $Q$ and $U$ maps and covariance matrices from the foreground fitting scheme acting on all input maps.   To allow for cross-checking, we also consider results from cleaned 100 GHz maps (i.e.\ from the foreground fitting with 143 GHz maps omitted) and from cleaned 143 GHz maps (i.e.\ from the foreground fitting with the 100 GHz maps omitted). 
\section{Tests on simulations} \label{sec:results_sims}
To investigate robustness, we test both likelihoods on realistic simulations with correlated noise. Therefore, we generate 100 Gaussian realisations of the CMB with an input $\tau=0.06$ and with (i) $r=0$ and (ii) $r=0.3$. We co-add these signal maps with realistic noise and systematics \SRtwo end-to-end (E2E) simulations\footnote{We remind the reader that for these testing purposes we use end-to-end simulations that were kept back from entering into the NCM fitting described in Sec.\ \ref{sec:NCM}.}. We perform a two-dimensional scan in $r$ and $\tau$. When scanning over $\tau$, we allow $A_s$ to change whilst keeping $10^9 A_s \exp{(-2 \tau)} =1.870$ fixed. For the two-dimensional scans we explore a broad parameter space by using a range of $\tau=0.01 - 0.1$ with $\Delta \tau = 0.005$ and $r=0.0 - 1.5$ with $\Delta r = 0.02$. The other parameters of the base $\Lambda$CDM cosmology are fixed to \Planck 2018 best-fit values \citepalias{Aghanim:2018eyx} when generating theoretical angular power spectra $C_{\ell}^{\rm th}$.

\begin{figure}
    \centering
     \includegraphics[width=\columnwidth]{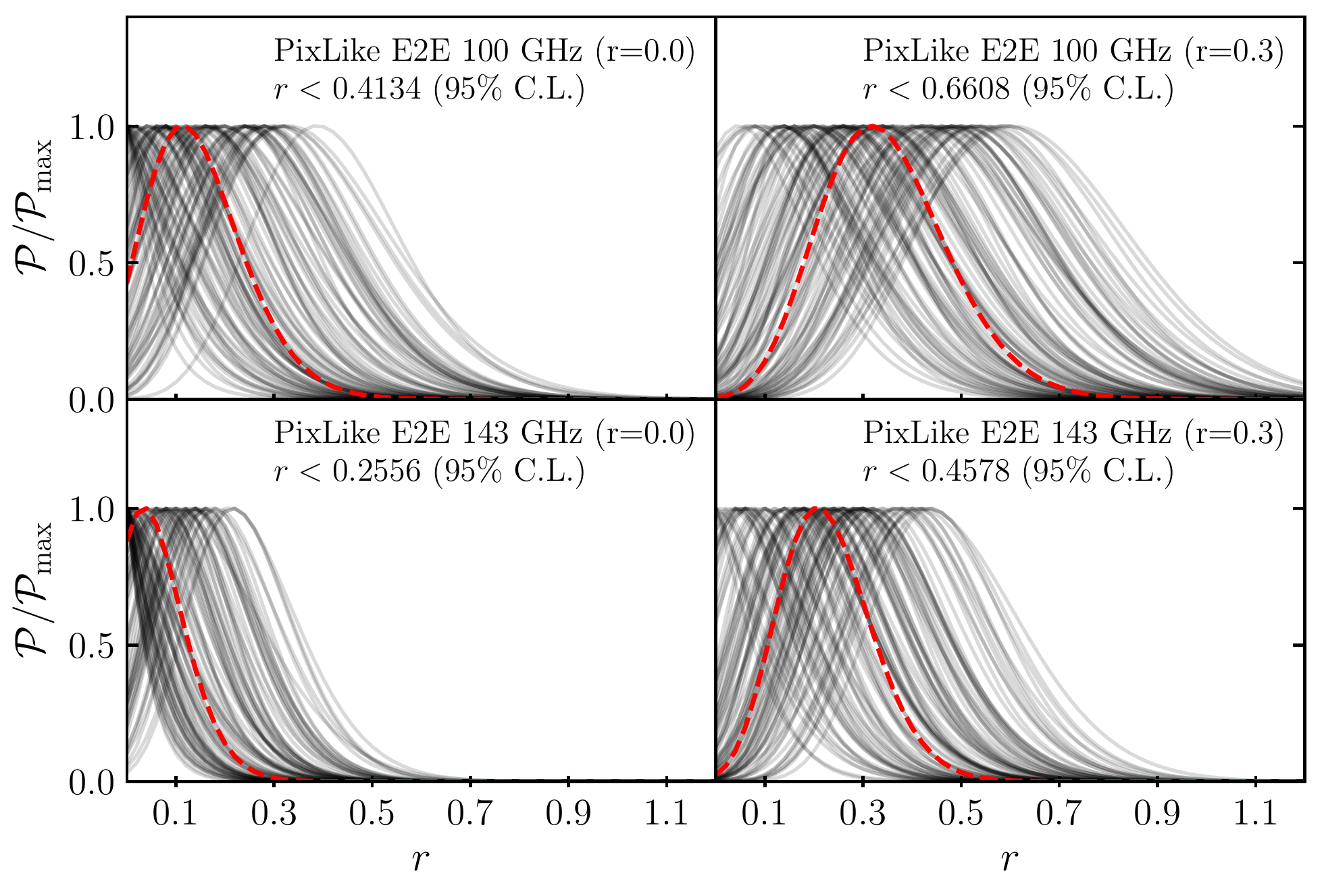}
    \vspace{-0.1in}
    \caption[]{One hundred posteriors, normalised to unity peak height, for $r$ (marginalised over $\tau$) from a two-dimensional scan on CMB realisations with realistic noise and systematics from \SRtwo end-to-end simulations obtained using our pixel-based likelihood (\pixlike). The left (right) column shows input CMB realisations with $r=0.0$ ($r=0.3$). The upper (bottom) row shows full-mission simulations using cleaned 100 GHz (143 GHz) maps. The red dashed curve is the geometric mean of the normalised posteriors.}
    \label{fig:simulation_test_pixlike}
    \vspace{-0.1in}
\end{figure}

\begin{figure}
    \centering
     \includegraphics[width=\columnwidth]{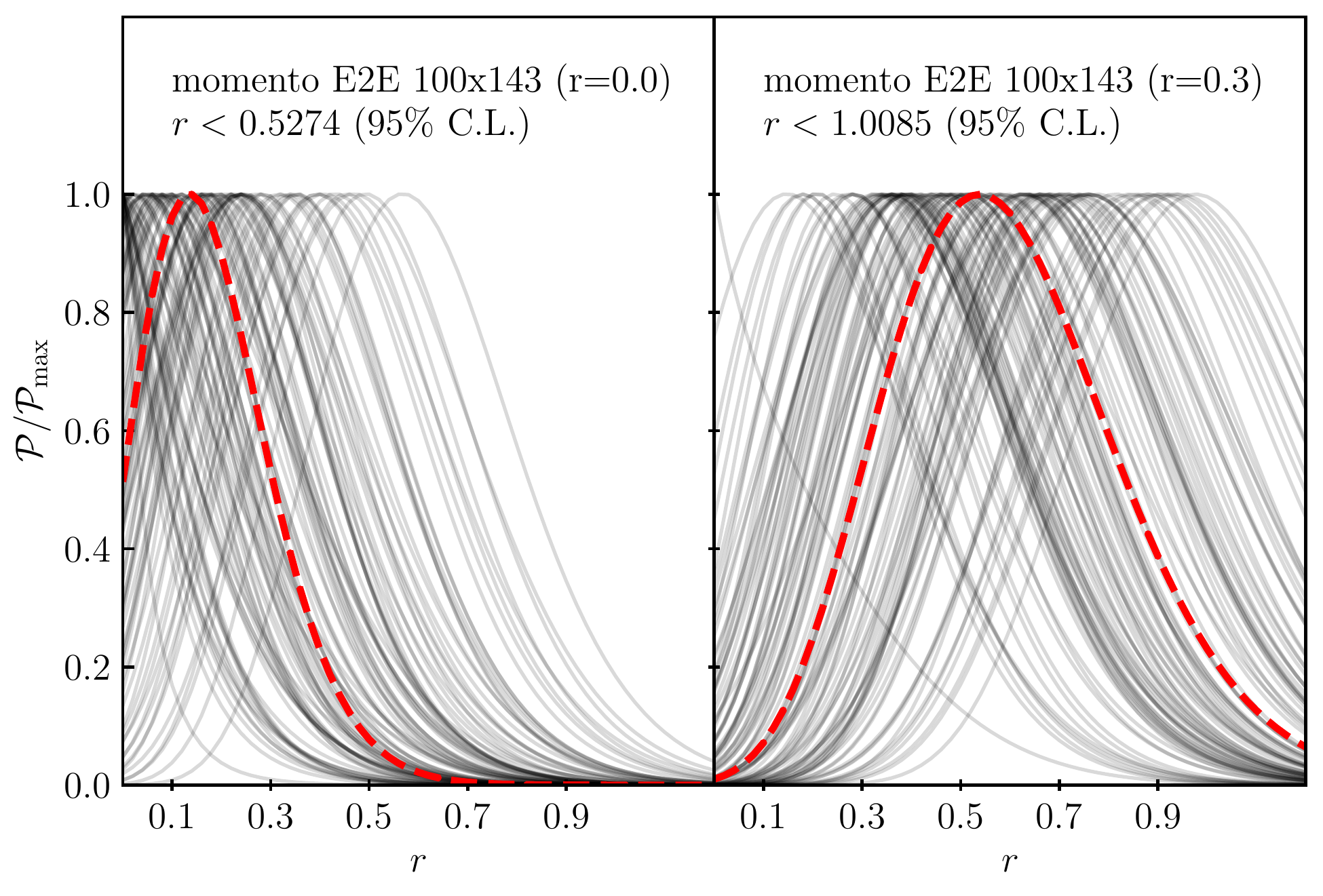}
    \vspace{-0.1in}
    \caption[]{One hundred posteriors, normalised to unity peak height, for $r$ (marginalised over $\tau$) from a two-dimensional scan using the \momento\ likelihood-approximation scheme.  The CMB realisations have been co-added with realistic noise and systematics from \SRtwo end-to-end simulations. We use 100$\times$143 full-mission cross-spectra with theory input values for $r=0$ (left panel) and $r=0.3$ (right panel). The red dashed curve is the geometric mean of the normalised posteriors.}
    \label{fig:simulation_test_momento}
    \vspace{-0.1in}
\end{figure}

Posteriors for $r$ after marginalizing over $\tau$, normalised to unity peak height, are shown in Fig.~\ref{fig:simulation_test_pixlike} for \pixlike and in Fig.~\ref{fig:simulation_test_momento} for \momento\footnote{We cross-check that the input $\tau$ values can be recovered with high fidelity when marginalising over $r$ and obtain matching results to figure 6 in \citetalias{Belsunce21}.}. Each solid black line represents the posterior of a simulation and the red dashed line shows the geometric mean of the normalised posteriors, given by $\Bar{\mathcal{P}}=(\prod_{i=1}^{N=100} (\mathcal{P}_i/\mathcal{P}_{i,\rm max}))^{1/N}$. The mean recovered upper limits \CL for both tests are quoted in each figure legend. Both methods perform qualitatively correctly for the given test, i.e.~ the input values of $r$ are recovered at the $2\sigma$ level and the mean upper limits on $r$ move upwards by $r\sim 0.2-0.5$ when increasing $r$ from $0$ to $0.3$. From the distributions shown in Figs.~\ref{fig:simulation_test_pixlike} and \ref{fig:simulation_test_momento}, \pixlike appears to return slightly tighter constraints, with the 143 GHz \pixlike likelihood appearing 
to be biased low and the \momento likelihood to be  biased slightly high.

We have looked in more detail for biases in our likelihoods by analysing the number of posteriors peaking at (or below) our input value of $r$ compared to the number peaking above that value\footnote{Assuming a binomial distribution, we expect a scatter of $\sqrt{N}/2$ in the distribution of posterior peaks around the input values of $r$, i.e.~for 100 simulations this corresponds to $5$ or $10$ peaks at the $1$ or $2\sigma$ level respectively.}. For \pixlike we obtain 61 and 48 posteriors that peak above zero\footnote{In the present analysis we explicitly do \emph{not} introduce $r_{\rm eff}$ as done in \citet{Tristram:2021a} as negative values for $r$ are unphysical.} for 100 and 143 GHz, respectively, for $r=0$. Increasing $r$ to 0.3 yields 46 and 24 posteriors that peak above $r=0.3$. Note that the spread in the posteriors is smaller for the 143 GHz channel than for the 100 GHz one. For \momento  we obtain 68 posteriors that peak above zero for $r=0$ and for $r$ of 0.3 leads to 72 posteriors peaking above the input value. The spread of the posteriors for \momento is significantly larger than for \pixlike. The mean upper limits of the posteriors of \momento are higher by $r\sim 0.2-0.5$ compared to the \pixlike ones.

The geometric means of the posteriors, shown as the red dashed curves, paint a similar picture. We evaluate the geometric means $\Bar{\mathcal{P}}$ at the input values of $r$. We obtain likelihood ratios from the peak corresponding to $1.3\sigma$ and $0.5\sigma$ deviations for the 100 and 143 GHz \pixlike likelihoods respectively. For $r$ of 0.3 we obtain $0.1\sigma$ and $0.9\sigma$ deviations, again, for 100 and 143 GHz respectively. For \momento, this test yields a difference of $1.2\sigma$ for $r=0$ and $1.1\sigma$ for $r=0.3$. This illustrates that the posteriors are behaving reasonably and are unbiased at the $\sim \sigma$ level.

To further understand if the noise modelling or the likelihood methods themselves may be responsible for these potential biases, we present additional tests in Appendix \ref{app:results_Gaussian_sims}.  These tests used Gaussian noise realisations generated directly from the assumed NCMs rather than noise realisations from end-to-end simulations.  Whilst these tests give us confidence that the likelihoods give unbiased results, they illustrate their sensitivity to the noise and systematics modelling captured in the covariance matrices. Both likelihoods perform reasonably well at the $2\sigma$ level.

The tests on simulations illustrate the challenges associated with constraining $r$ from low-$\ell$ \Planck data alone. The spread in the distribution of posteriors show the need for high precision polarisation cross-spectrum measurements. These, in turn, depend sensitively on the signal-to-noise ratio in the data and on the handling of systematics in the noise description. Because of the sensitivity of our methods to the noise modelling, the tests described here suggest that our results on $r$ should be interpreted as indicative rather than as rigorous 95\% confidence limits. 

\section{Results} \label{sec:results}
In this section, we present results for the tensor-to-scalar ratio $r$ using the likelihoods \pixlike and \momento, discussed above. To quantify the \Planck constraints coming from low-multipole polarisation data and to reduce the computational burden, we perform two-dimensional scans in $\tau$ and $r$ rather than combining with a high-multipole likelihood and scanning 
over the full set of $\Lambda$CDM parameters (which would return constraints on $r$ dominated by the fit to the temperature power spectrum). Throughout the paper, we fix other cosmological parameters to those of the \Planck best-fit base $\Lambda$CDM model \citepalias{Aghanim:2018eyx}: $H_0 = 67.04$, $\Omega_{\rm b}h^2 = 0.0221$, $\Omega_{\rm c}h^2 = 0.12$, $\Omega_{\nu}h^2 = 0.00064$, $\theta_*=1.0411$, $n_{\rm s}=0.96$ and vary $A_s$ as a function of $\tau$ to keep $A_s e^{-2\tau} = 1.870$ fixed.

\begin{figure}
    \centering
     \includegraphics[width=\columnwidth]{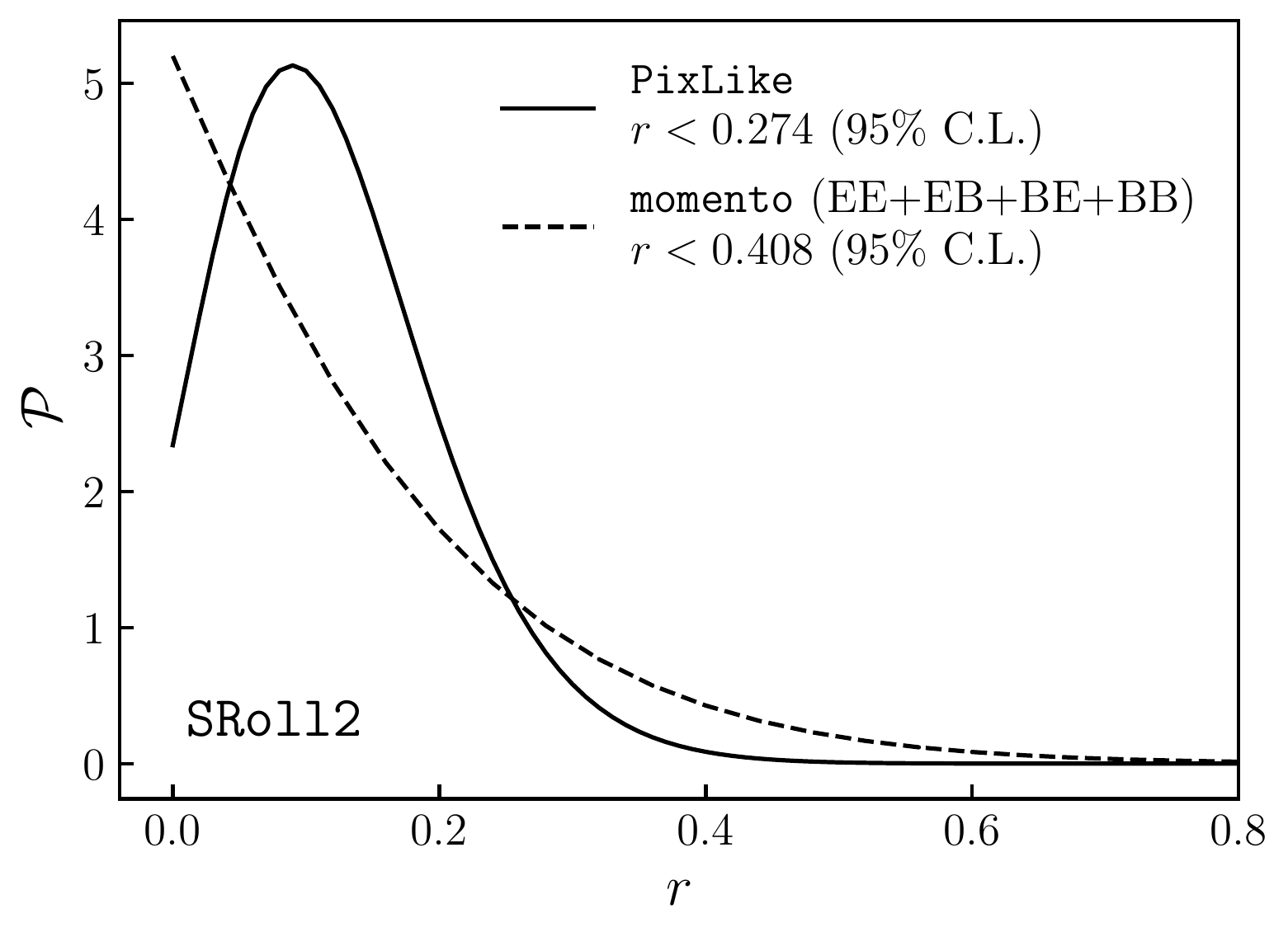}
    \vspace{-0.2in}
    \caption[]{Posteriors for $r$ (marginalised over $\tau$) from  two-dimensional low-$\ell$ scans, using a pixel-based likelihood (\pixlike) and a QCS-based likelihood approximation scheme (\momento), on NR-cleaned polarisation \Planck maps. The uncertainties from the foreground cleaning have been propagated through to the final NCMs.}
    \label{fig:2D_posterior}
    \vspace{-0.1in}
\end{figure}

For our quoted $r$ constraints, we use NR-foreground-cleaned-maps with \Planck 2018 \madam data for LFI (30 and 70 GHz) and \SRtwo data for HFI (100, 143, 217, and 353 GHz) with the corresponding covariance matrices. \pixlike uses full-mission polarisation $Q$ and $U$ maps, whereas \momento uses half-mission polarisation maps. For \momento\ the cleaning procedure is applied twice, once using polarisation maps from the first half of the mission (HM1) and then using maps from the second half of the mission (HM2). \momento then computes HM1$\times$HM2 $EE$, $EB$, $BE$, and $BB$ quadratic cross spectra to constrain $r$. We find the following upper limits at 95\% confidence limit (C.L.):
\begin{subequations}
\begin{align}
r^{\pixlike}_{\kpivot}  &< 0.274\ \quad &&(\text{\Planck low-}\ell\ Q,\ U\ \text{maps}),   \label{eq:r1} \\
r^{\momento}_{\kpivot}  &< 0.408 \quad &&(\text{\Planck low-}\ell\ EE,\ EB,\ BE,\ BB),   \label{eq:r2}
\end{align}
\end{subequations}
which are both consistent with $r=0$. The shapes of the posteriors are shown in Fig.\ \ref{fig:2D_posterior} with both of these limits being very substantially weaker than the \cite{Tristram:2021b} constraints of Eq.~\eqref{eq:Tristram_r}.  The pixel-based likelihood peaks around $r=0.1$ and yields a tighter constraint than \momento. This comes primarily from the fact that the pixel-based likelihood is in principle optimal, using information from the full-mission maps. So in comparison to \momento, \pixlike effectively uses auto-spectra in addition to cross-spectra and does not lose any other information in the compression from low-resolution maps to QCS, leading to a theoretical $\sqrt{2}$ improvement in signal-to-noise in the noise-dominated regime. 

\begin{figure*}
    \centering
    \includegraphics[width=\textwidth]{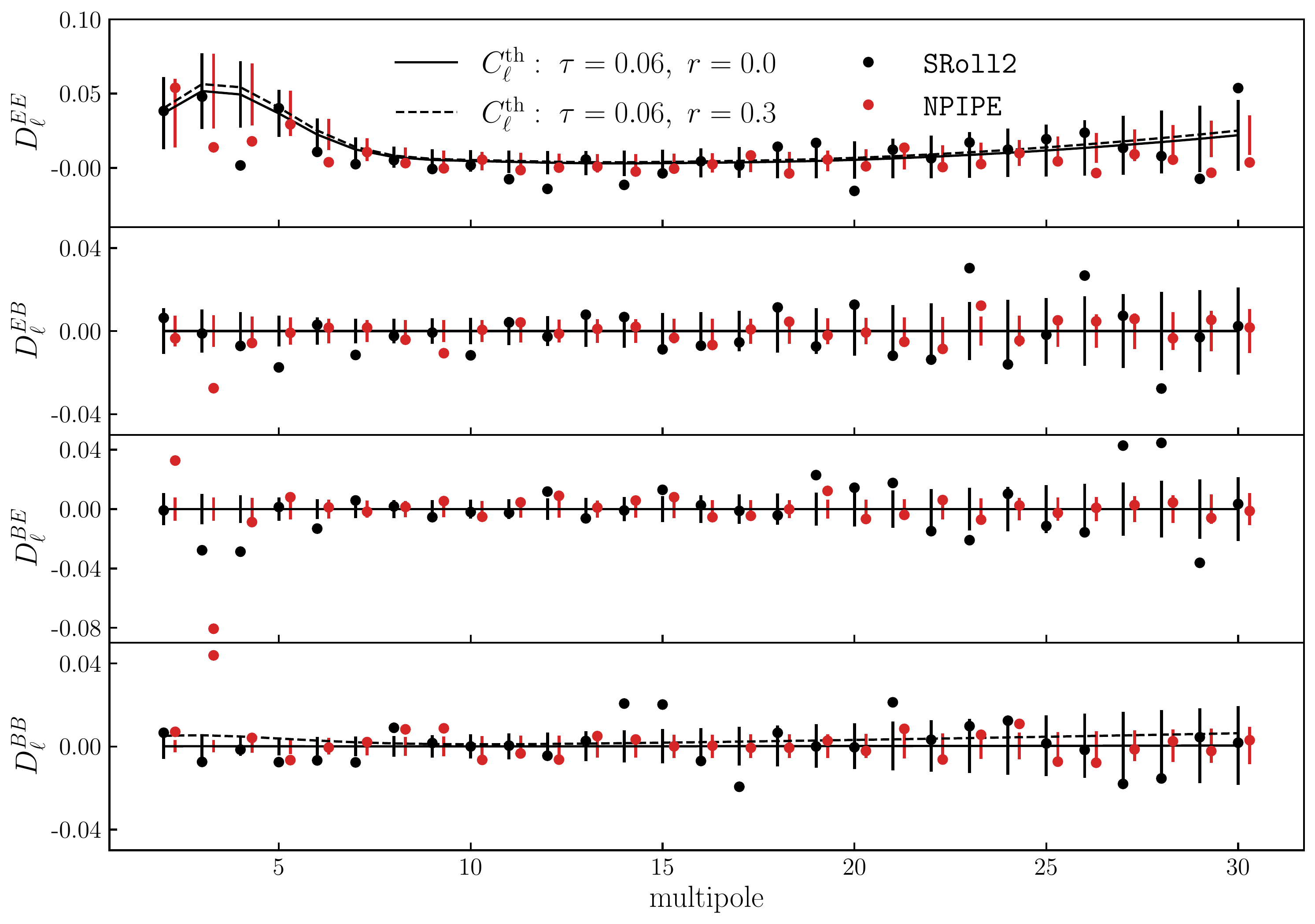}
    \vspace{-0.2in}
    \caption[Power spectrum from \SRtwo data]{$EE$, $EB$, $BE$ and $BB$ polarisation QCS for NR-cleaned HM1$\times$HM2 \SRtwo (black) and  $100\times 143$ full-mission template-cleaned \NPIPE (red) data. A theoretical angular power spectrum is shown in black with best-fit $\Lambda$CDM parameters and two parameter combinations: (i) $\tau=0.06$ and $r=0$ (solid); and (ii) $\tau=0.06$ and $r=0.3$ (dashed). The error bars (square roots of the diagonals of the covariance matrices) are plotted on the theory curve with $r=0$.}
    \label{fig:momento_QCS_r}
    \vspace{-0.1in}
\end{figure*}

The NR-cleaned HM1$\times$HM2 $EE$, $EB$, $BE$ and $BB$ cross spectra used in \momento are shown in Fig.\ \ref{fig:momento_QCS_r}.  The solid linet shows a theoretical model with $\tau = 0.06$ and $r=0$ and to guide the eye,
the dashed line shows the same model but with $r=0.3$ (close to the 95\% C.L. upper limit of Eqs.~\eqref{eq:r1} and \eqref{eq:r2}). The tensor power at the reionization bump scales as $r\tau^2$. The $EE$ spectrum constrains $\tau$ and given this the constraint on $r$ then comes almost entirely from the $BB$ spectrum. The $EB$ and $BE$ spectra provide a consistency check since in the absence of parity violating physics, these spectra should be identically zero. Indeed, eliminating the $EB$ and $BE$ spectra from \momento we find:
\begin{align}
& r^{\momento}_{\kpivot}  < 0.409 \qquad (\text{\Planck low-}\ell\ EE,\ BB).   \label{eq:r3}
\end{align}
Further eliminating the $EE$ spectrum makes the $r$ constraint sensitively dependent on the (lower limit of the) $\tau$ prior (as $\tau$ is no longer constrained). For example, with a uniform $0.04 \leq \tau \leq 0.1$ prior we find
\begin{align}
& r^{\momento}_{\kpivot}  < 0.439 \qquad (\text{\Planck low-}\ell\ BB,\  \tau\, \mathrm{prior\, as\, above}).   \label{eq:r4}
\end{align}
The $BB$ spectrum plotted in Fig.\ \ref{fig:momento_QCS_r} is consistent with zero, though the negative points at $\ell = 5-7$ may indicate  residual systematic  biases in  the \Planck\ spectra. This run of negative points would cause a \momento posterior for $r$ using $\ell=2-7$ to peak even more strongly around $r=0$ than the one plotted in Fig.~\ref {fig:2D_posterior}. The latter posterior, however, includes high points at $\ell=8,14$ and 15 that favour positive values of $r$, partially compensating the downward pull.

For our constraints on the optical depth to reionization $\tau$ in \citetalias{Belsunce22}, we obtained consistent results for \momento and \pixlike from NR-foreground cleaned \Planck maps. Together with the rough agreement seen here for $r$ also, this gives us some confidence that the NCMs and smoothed templates we have constructed are capturing noise and systematic effects well enough for outputs from the pixel-based scheme to be of interest, at least when using \SRtwo\ products. 

\subsection{Comparison to \NPIPE results} \label{sec:NPIPE_results}
In the following we apply our low-multipole polarised CMB noise fitting and inference pipeline to the \NPIPE maps \citep{NPIPE:2020}. The \NPIPE maps use slightly more time-ordered data than the maps
produced for the  \Planck  2018 data release and they  apply different
algorithms to clean and calibrate the time-ordered data. In addition,  the \NPIPE map making procedure applies a  polarisation prior,  effectively suppressing power at the largest angular scales ($\ell <10$); see \citet{NPIPE:2020} for details. The CMB maps thus have to be corrected by transfer functions measured from end-to-end simulations. These transfer functions are defined as the ratio of the estimated to the theoretical input power spectrum in harmonic space. For an exact treatment, the maps have to be deconvolved by the transfer functions which depend on multipoles $\ell$ and varying cosmology, i.e.~$r$ and $\tau$. For the present analysis, however, we follow the approach used in \citet{Tristram:2021b} and apply the transfer function provided by \NPIPE. These have been computed using a sky mask with a value of $f_{\rm sky}=0.6$\footnote{We note that our sky mask uses $f_{\rm sky}\approx 0.54$.}, and we analogously apply the same $EE$ transfer function to $EB$, $BE$, and $BB$ (see appendix A in \citet{Tristram:2021b}). While this is clearly sub-optimal, it allows for an approximate comparison of the \SRtwo and \NPIPE noise levels and illustrates the challenges associated with using \NPIPE for inference at the lowest multipoles ($\ell < 10$)\footnote{We emphasise that while the present treatment of the transfer function is approximate, we do not perform the NR-cleaning procedure on \NPIPE maps. This treatment suffices for a cross-spectrum-based comparison of our \SRtwo-based analysis to \NPIPE-based ones \citep[presented in e.g.~][]{NPIPE:2020,Tristram:2021a,Tristram:2021b, Beck:2022}.}. 

For the present \NPIPE analysis, we construct NCMs using the noise fitting procedure presented in Sec.~\ref{sec:NCM}. As base matrix we use the \NPIPE low-resolution map-making covariance matrix, $\mat{N}_0$, and add large-scale modes into $\mat{N}_0$. Using Eq.~\eqref{eq:likelihood_ncm} we fit the base matrix to the covariance matrix computed from \NPIPE end-to-end simulations. This procedure is similar to the modelling of NCMs from the \SRtwo simulations, but for  \NPIPE\ we find lower noise levels at low multipoles compared to
 \SRtwo resulting in tighter error bars for the \NPIPE\ quadratic cross-spectra by a factor of 1.5 - 3. The QCS measurements for \NPIPE full-mission $100\times143$ GHz data are shown in Fig.~\ref{fig:momento_QCS_r} in black. 

\begin{figure}
    \centering
     \includegraphics[width=\columnwidth]{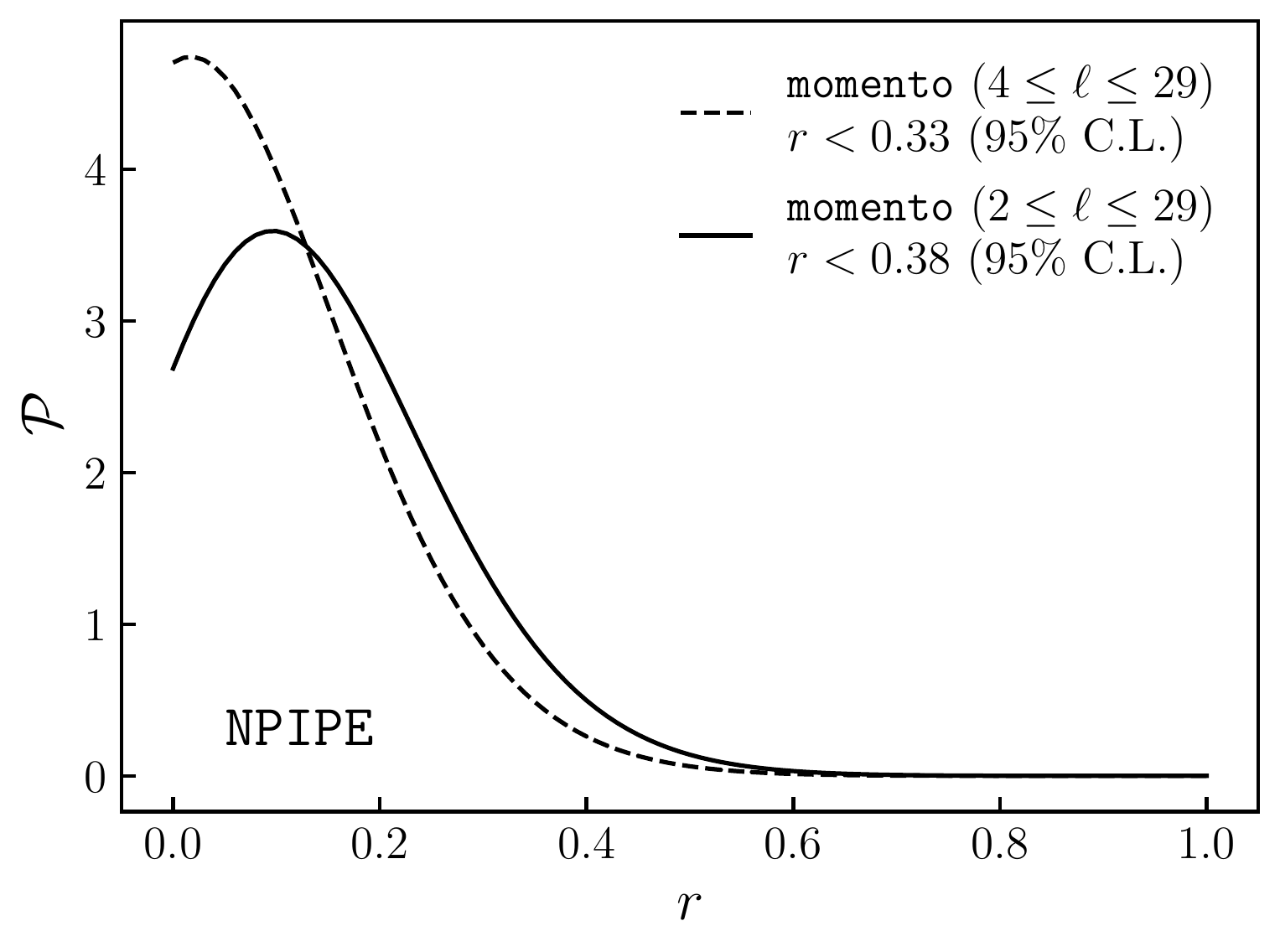}
    \vspace{-0.2in}
    \caption[]{Posteriors for $r$ (marginalised over $\tau$) from  two-dimensional low-$\ell$ scans, using a cross-spectrum-based likelihood (\momento), on template-cleaned $100\times143$ full-mission \NPIPE maps. \momento uses all the \NPIPE polarisation spectra ($EE$, $EB$, $BE$ and $BB$). The dashed posterior uses the full multipole range ($2\leq \ell \leq 29$) and the solid posterior removes the first two multipoles ($4\leq \ell \leq 29)$ of all spectra since these are quite very noisy and most affected by the transfer function.}
    \label{fig:2D_posterior_NPIPE}
    \vspace{-0.1in}
\end{figure}

In general, the QCS estimates for \NPIPE are less noisy than for \SRtwo, especially at multipoles $\ell >10$, i.e.~where the effect of the transfer function is below 10\%\footnote{Table G.1 in \citet{NPIPE:2020} lists the relevant values for the transfer functions $\ell$-by-$\ell$ per frequency channels. In the present analysis, we apply the $100\times143$ GHz transfer function measured on end-to-end simulations.}. At the lowest multipoles $2 \leq \ell \leq 4$, however, several strong outliers appear in all polarisation spectra. To constrain cosmological parameters, we perform a two-dimensional scan over $\tau$ and $r$ using all the available \NPIPE polarisation spectra ($EE$, $EB$, $BE$ and $BB$) for $100\times 143$ full-mission data, yielding a 95\% confidence limit for the tensor-to-scalar ratio of
\begin{equation}
r^{\momento}_{\kpivot}  < 0.38 \quad (\text{\NPIPE low-}\ell\, EE,\ EB,\ BE,\ BB),   \label{eq:r5}
\end{equation}
which is shown in Fig.~\ref{fig:2D_posterior_NPIPE}. To assess the impact of the \emph{very} strong $\ell=3$ outlier in the $BB$ cross-spectrum\footnote{\citet{Tristram:2021b} find a similar outlier at $\ell=3$, even for varying $f_{\rm sky}$.}, we also compute $r$ using the multipole range $4\leq \ell \leq 29$ and obtain, as expected, a tighter result of $r<0.33$ \CL. 

We do not apply the Bayesian component separation method to \NPIPE frequency maps, discussed in Sec.~\ref{sec:data}, given that we do not have the required transfer functions. We did, however, compute constraints on $r$ using the pixel-based likelihood using template-cleaned \NPIPE frequency maps at 100 GHz and at 143 GHz. These both returned posteriors very tightly peaked at zero ($r\simlt 0.12, 0.11$ at 95\% C.L. respectively). This behaviour left us concerned that our noise covariance matrices and transfer-function-handling are not accurate enough for constraining $r$ using our pixel-based scheme with \NPIPE template-cleaned maps. See App.~\ref{app:reg_noise} for a further discussion of the sensitivity of \pixlike to the NCMs and regularising noise.

To cross-check our results, we compute constraints for the optical depth to reionization $\tau$ and obtain, using the same \NPIPE polarisation cross-spectra, a value of 
\begin{equation}
    \tau = 0.055 \pm 0.0056 \quad (\text{\NPIPE low-}\ell\, EE,\ EB,\ BE,\ BB).   \label{eq:tau_NPIPE}
\end{equation}
Our result is in agreement with the value quoted in \citet{NPIPE:2020} obtained using only the $EE$ cross-spectrum, albeit $\sim 0.5 \sigma$ higher, which gives us some confidence in the computed QCS estimates, the noise fitting procedure, and the likelihood in this situation. 

\section{Summary and Conclusion} \label{sec:conclusion}
Measuring primordial $B$-modes, parametrised by the tensor-to-scalar $r$, from large-angular scale ($2 \leq \ell \leq 30$) polarised CMB data is challenging, given the low amplitude of the signal in comparison to Galactic polarised foregrounds and potential sources of systematic error. The determination of $r$ is a key science goal of upcoming CMB experiments and crucial to our understanding of inflation and the dynamics involved. This paper has coupled the significantly improved mapmaking algorithms for HFI \citep{2019A&A...629A..38D} with mathematically well-motivated techniques for foreground cleaning and likelihood construction. We have used maps that have been cleaned using a Bayesian parametric foreground fitting model \citepalias{Belsunce22}, and propagated the uncertainties from the cleaning through to the final constraints on $r$.
We have compared two likelihoods: (i) a polarisation-only pixel-based likelihood (\pixlike), formally exact even on a masked sky, and (ii) a likelihood-approximation scheme (\momento) that uses all the available polarisation cross-spectra, namely $EE$, $EB$, $BE$ and $BB$, to effectively construct 
a least presumptive sampling distribution \citep{Gratton:2017}. The corresponding constraints are given in Eqs.\ \eqref{eq:r1} and \eqref{eq:r2}. The pixel-based likelihood yields tighter results by $\sim 30\%$ because it uses more data than our implementation of \momento\ which uses only cross-spectra for robustness considerations. It is reassuring that the pixel-based likelihood gives consistent results given our simplified models for the noise covariance matrices and the removal of systematic template maps derived from end-to-end simulations. Additionally, we extend our cross-spectrum-based likelihood to full-mission $100\times143$ GHz \NPIPE data, quoted in Eq.~\eqref{eq:r5}, and obtain similar constraints as for \SRtwo. 

Our main conclusion is that both likelihoods are in agreement with each other but \emph{do not} support the claims of very tight constraints on $r$ from \Planck data alone made in \citet{Tristram:2021a, Tristram:2021b}. This is an important result because at present \Planck is the only CMB experiment that can probe low multipoles in polarisation. The likelihood approximation used by \citet{Tristram:2021a,Tristram:2021b} 
is heuristic and has difficulties in handling cross-spectral multipoles that happen to be negative, and their results are unusually sensitive to the multipole range $\ell \sim 30-40$. 
In addition, a reanalysis of the above results by \citet{Beck:2022} using `conditioned' matrices, where effectively off-diagonal terms are set to zero, substantially broadens the constraints from \citet{Tristram:2021a,Tristram:2021b} for $r$ on \NPIPE data, leading to results consistent with this paper\footnote{Note also that for a distribution given by restricting and renormalising a negatively-peaked Gaussian to positive values only, its standard deviation tends to $\sigma / | \mu |  $ of that of the parent Gaussian, with confidence limits scaling downwards similarly.  So a distribution that is pulled unphysically low inevitably leads to tight constraints.}. Our results agree with what we would expect from estimates of the noise levels in the \Planck data.

Given the noise levels and residual systematics in the \Planck polarisation maps it is unlikely that the large angular constraints on $r$ from \Planck can be improved with further data processing. Tighter limits on the optical depth to reionization $\tau$ and large angle $B$-modes will therefore require a new space mission. The CMB satellite LiteBIRD \citep{2020JLTP..199.1107S} will be optimised for such measurements and will require careful systematics control, likelihood construction and cosmological data analysis. 

\section*{Acknowledgements}
The authors thank the \Planck Collaboration and the Bware team for their tremendous efforts to produce wonderful data sets. Further the authors thank R. Keskitalo for useful discussions regarding \NPIPE data products. RdB acknowledges support from the Isaac Newton Studentship, Science and Technology Facilities Council (STFC) and Wolfson College, Cambridge. SG acknowledges the award of a Kavli Institute Fellowship at KICC. We acknowledge the use of: \texttt{HEALPix} \citep{2005ApJ...622..759G} and \texttt{CAMB} (\url{http://camb.info}). This work was performed using the Cambridge Service for Data Driven Discovery (CSD3), part of which is operated by the University of Cambridge Research Computing on behalf of the STFC DiRAC HPC Facility (\url{www.dirac.ac.uk}). The DiRAC component of CSD3 was funded by BEIS capital funding via STFC capital grants ST/P002307/1 and ST/R002452/1 and STFC operations grant ST/R00689X/1. DiRAC is part of the National e-Infrastructure.

\section*{Data availability}
The data underlying this article will be shared on reasonable request. The \Planck and \SRtwo maps and end-to-end simulations are publicly available.



\bibliographystyle{mnras}
\bibliography{references} 


\appendix
\section{Effect of regularising noise} \label{app:reg_noise}
\begin{figure}
    \centering
     \includegraphics[width=\columnwidth]{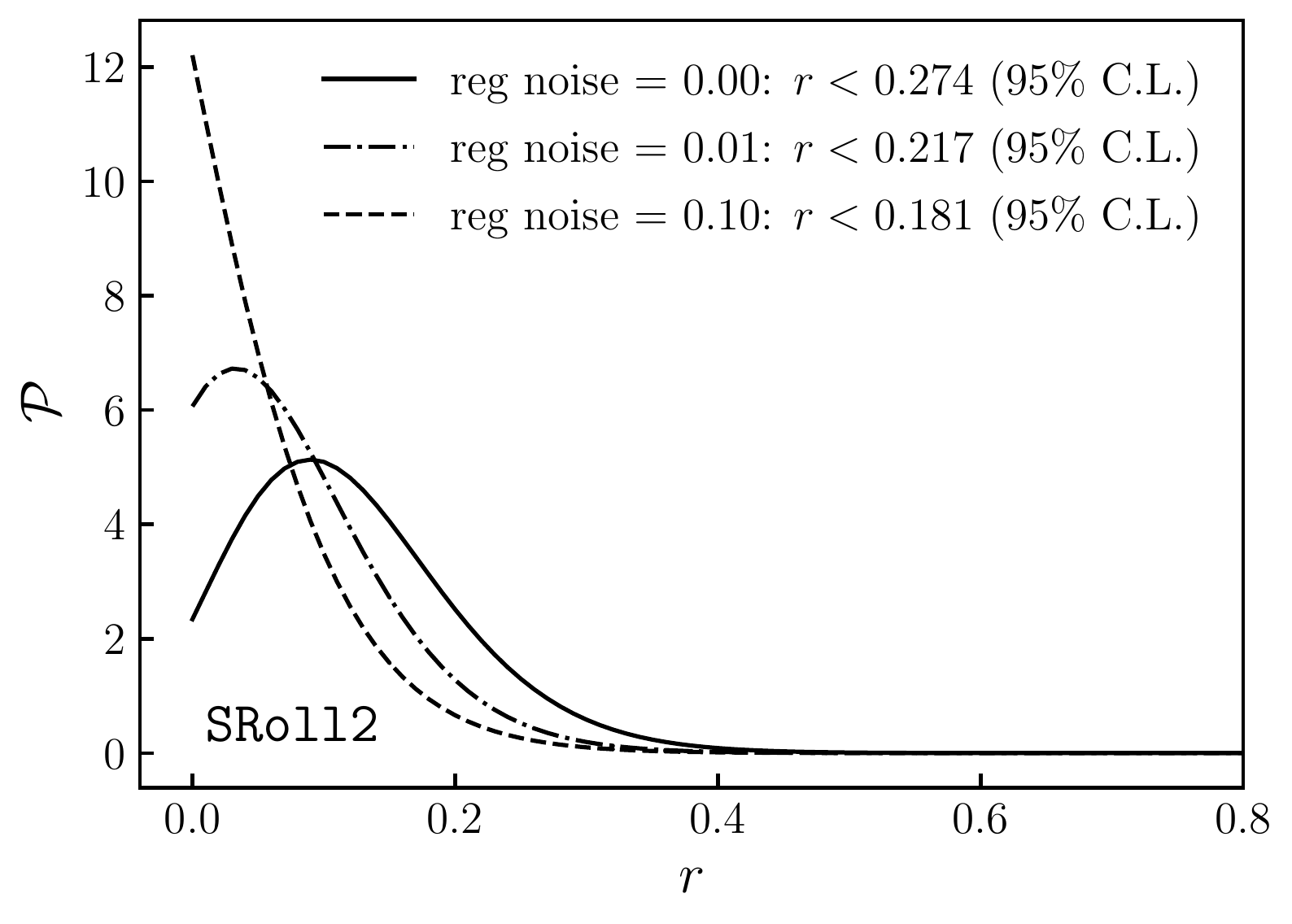}
    \vspace{-0.2in}
    \caption[]{Marginalised posteriors for $r$ over $\tau$ from a two-dimensional low-$\ell$ scan using the pixel-based likelihood (\pixlike) for \Planck polarisation data. The data has been cleaned using the NR method. The results for $r$ are shown as a function of regularising noise ``reg noise" (in units of $\mu \mathrm{K}$) that is squared and added to the diagonals of the NCMs. The solid curve is our best estimate for $r$ without any additional regularising noise added. The dashed (reg noise = 0.01) and dash-dotted (reg noise = 0.1) lines illustrate two levels of varying regularising noise. The area under the posteriors is normalised to unity. The upper limit on $r$ is quoted with a 95\% confidence limit (C.L.). (Note that the corresponding  $\tau$ constraints found after marginalising over $r$ are indistinguishable.)}
    \label{fig:2D_posterior_regnoise}
    \vspace{-0.1in}
\end{figure}

Regularising noise may be added to NCMs for two purposes in CMB analyses.  The first is as a technical operation simply to render NCMs invertible (typically one uses smoothed maps to prevent aliasing and the corresponding smoothed covariance matrices are then necessarily rank deficient). The second is to render a likelihood insensitive to the form of the theory spectrum above a certain scale.  This is desired when combining a low-$\ell$ likelihood with a high-$\ell$ one, stopping differences between the theory and the data being too highly penalised in the overlap region.  Depending upon the expected signal, the second application typically requires significantly more noise added than the first, and to avoid biases also needs an additional compensating term in the numerator of the likelihood \citep{GrattonPC}.

$B$-modes have small amplitudes compared to the magnitude of other present signals and foregrounds. In Fig.~\ref{fig:2D_posterior_regnoise} we illustrate the effect of adding varying amounts of regularising noise to the NCMs (without any addition of a compensating term). We note that increasing the regularising noise from 0.0 to $0.1\ \mu K$ pushes the upper limits on $r$ closer to zero and yields artificially tight constraints on $r$.  As we are not coupling the low-$l$ likelihood to a high-$l$ for the $B$-mode-focused $r$ analysis presented here, and in contrast with our $E$-mode-focused $\tau$ analysis of \citetalias{Belsunce21}, we need not be concerned about overlap effects.  Hence we proceed with minimal amounts of regularizing noise. 

These findings illustrate the importance of accurate and well-tested NCMs for parameter inference, in a complementary manner to the findings of \citet{Beck:2022}. 

\section{Tests on Gaussian realisations of the NCMs} \label{app:results_Gaussian_sims}
\begin{figure*}
    \centering
    \begin{subfigure}[b]{\columnwidth}
        \centering
        \includegraphics[width=\columnwidth]{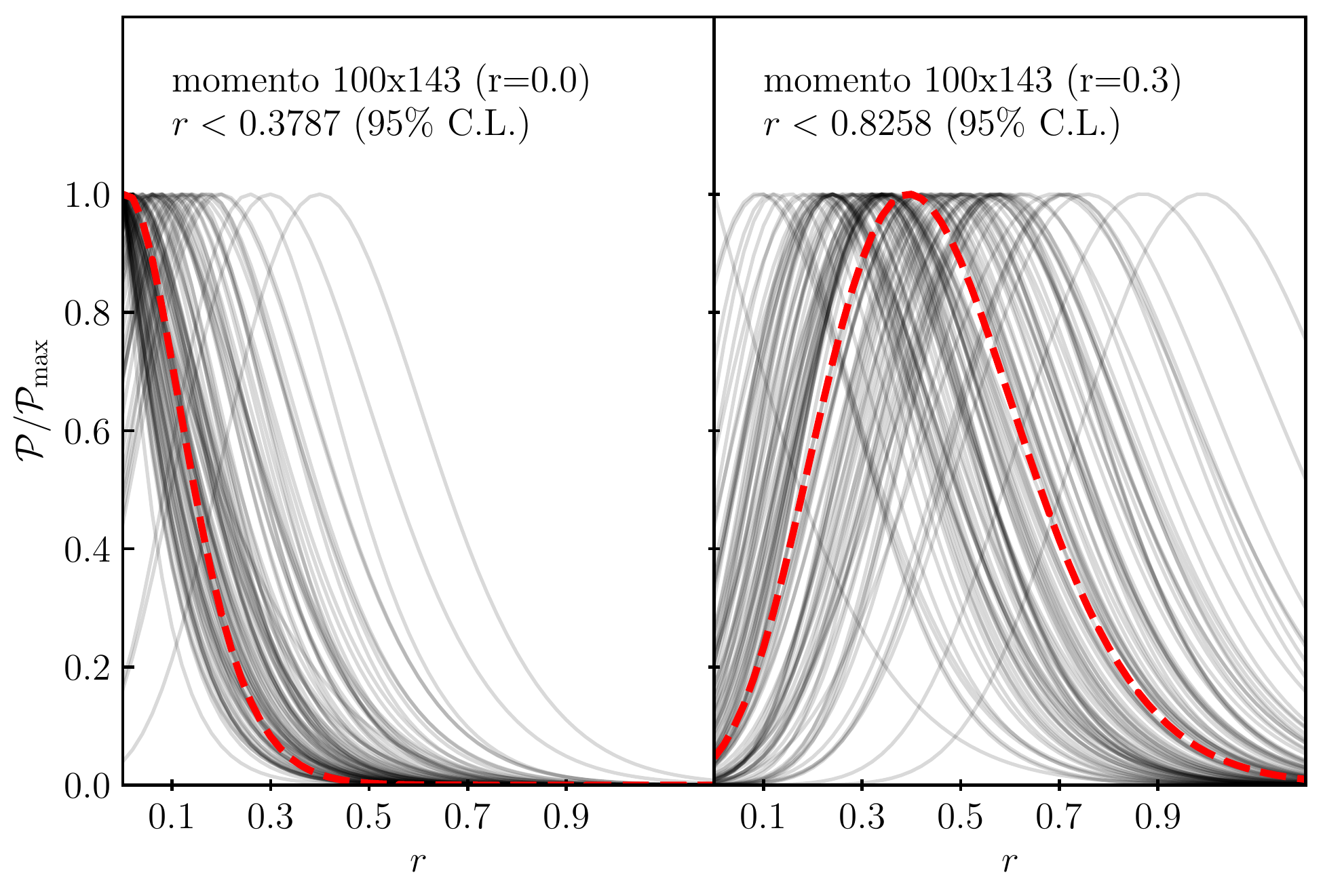}
        \caption{$100\times 143$ full-mission NCM.}
        \label{fig:simulation_test_momento_Gaussian_full}
    \end{subfigure}
    \hfill
    \begin{subfigure}[b]{\columnwidth}
        \centering 
        \includegraphics[width=\columnwidth]{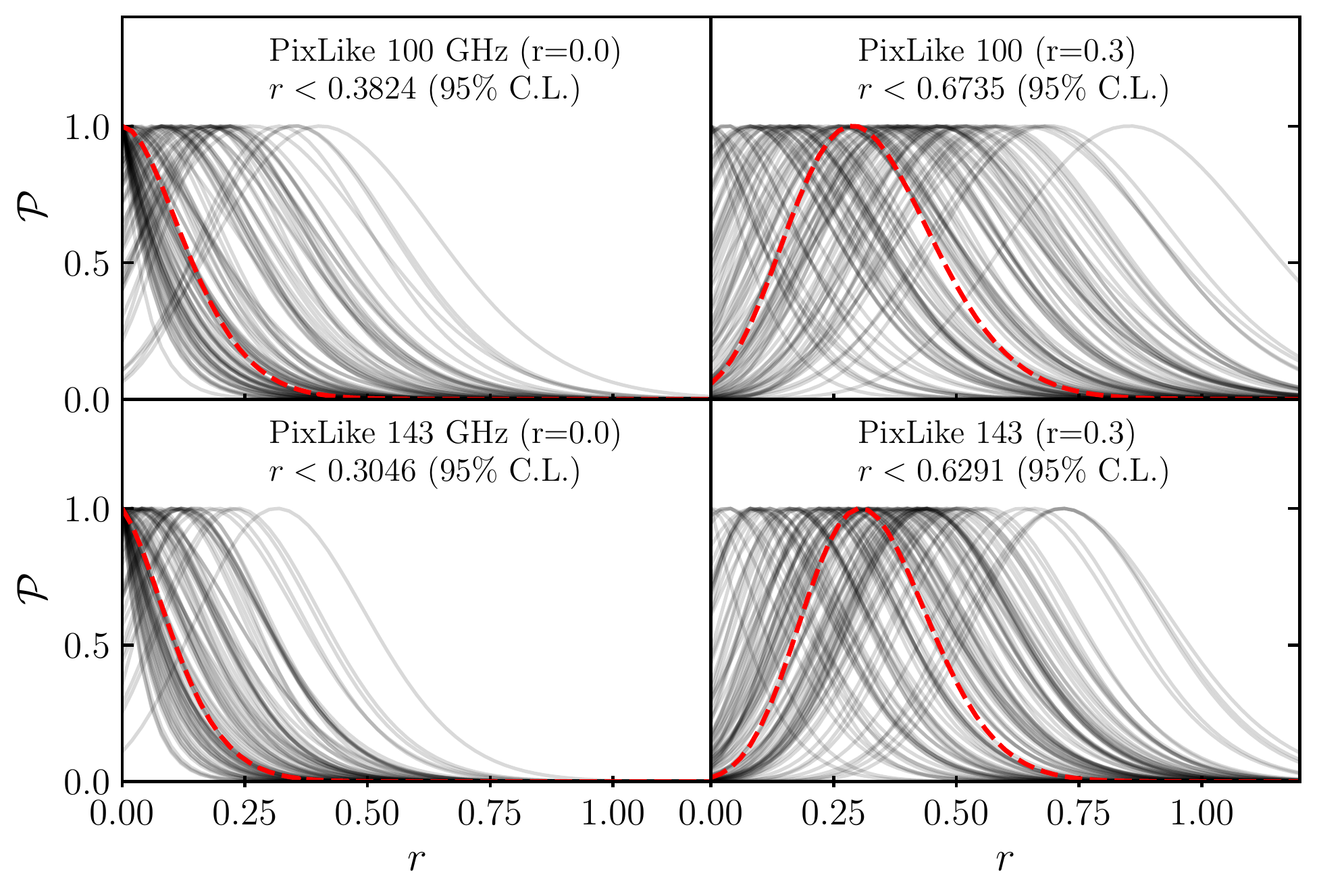}
        \caption{$100$ and $143$ full-mission NCM.}
        \label{fig:simulation_test_pixlike_Gaussian_full}
    \end{subfigure}
    \vskip\baselineskip
    \begin{subfigure}[b]{\columnwidth}
        \centering 
        \includegraphics[width=\columnwidth]{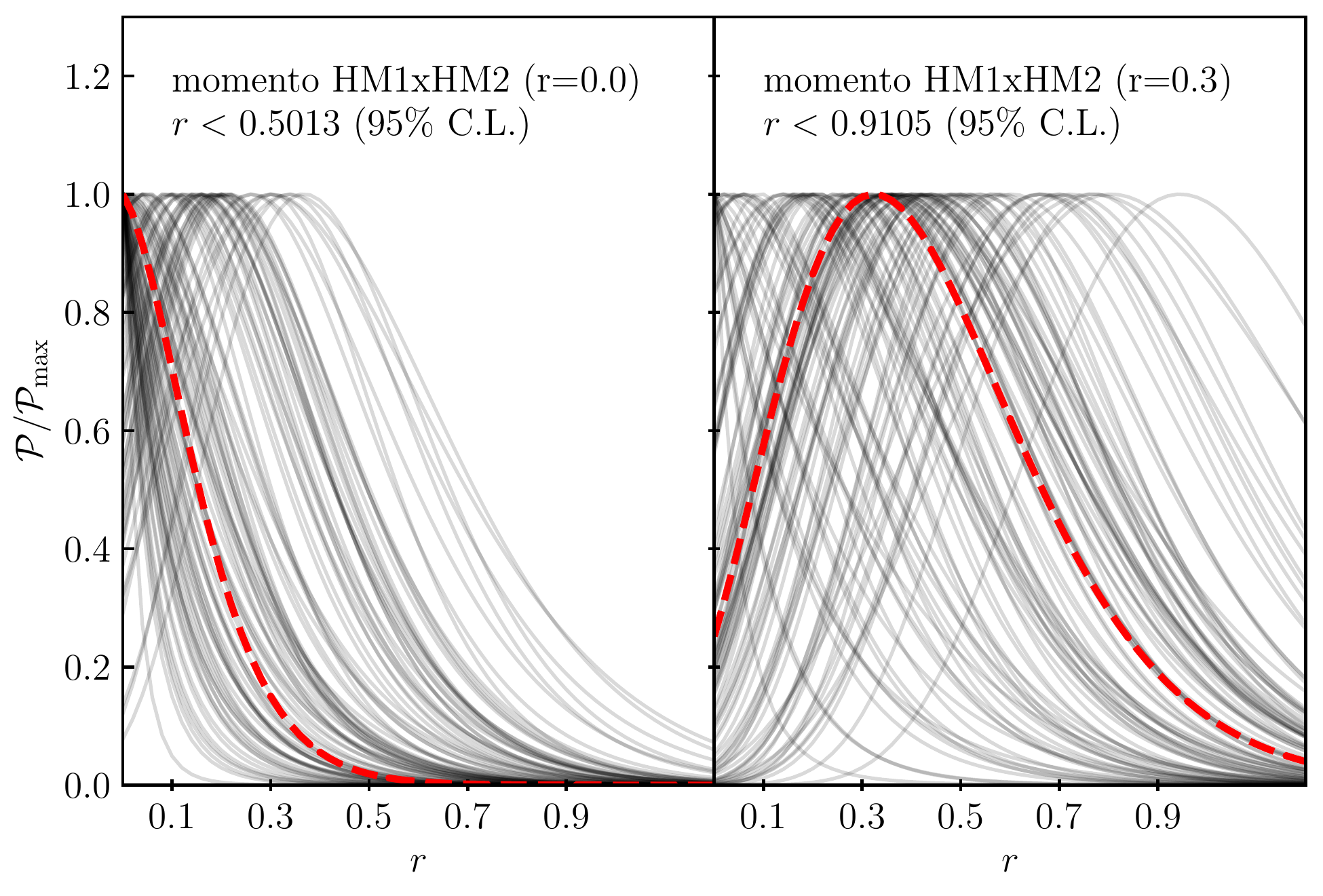}
        \caption{ HM1$\times$HM2 NCM from NR foreground cleaning.}
        \label{fig:simulation_test_momento_Gaussian_HM}
    \end{subfigure}
    \hfill
    \begin{subfigure}[b]{\columnwidth}
        \centering 
        \includegraphics[width=\columnwidth]{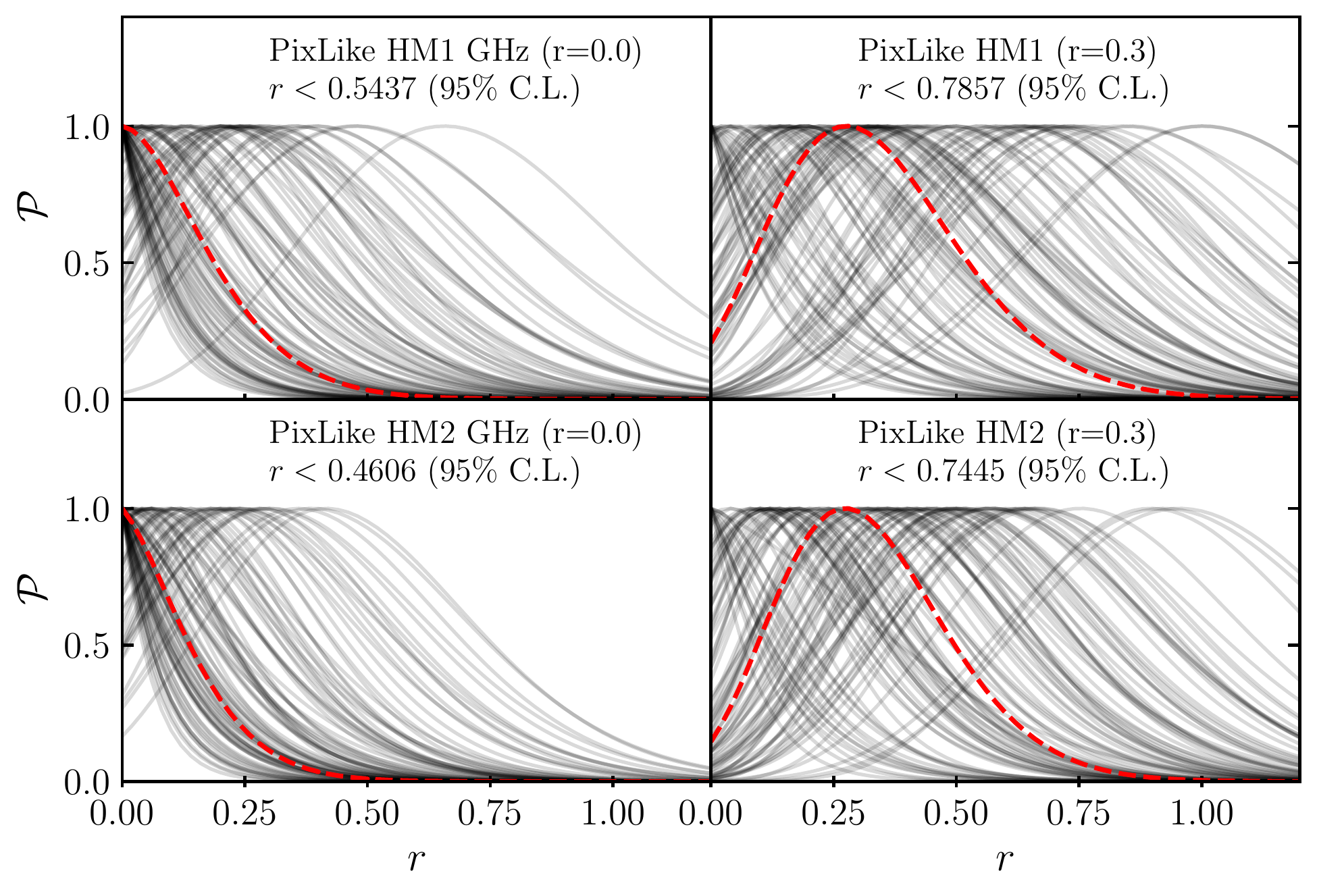}
        \caption{HM1 and HM2 NCM from NR foreground cleaning.}
        \label{fig:simulation_test_pixlike_Gaussian_HM}
    \end{subfigure}
    \caption{Sets of 100 posteriors, normalised to unity peak height, for $r$ (marginalised over $\tau$) from two-dimensional scans on Gaussian realisations of the CMB co-added with Gaussian realisations of the appropriate NCMs. These are shown for \momento (left panels) and  \pixlike (right panels), with input tensor-to-scalar ratios of $r=0$ (left subpanels) and $r=0.3$ (right subpanels). The top row uses $100\times 143$ full-mission NCMs (without any foreground cleaning uncertainties) and the bottom row uses half-mission NCMs (which include uncertainties from the foreground cleaning procedure). The red dashed curves are the geometric mean of the normalised posteriors.}
    \label{fig:simulation_test_Gaussian}
    \vspace{-0.1in}
\end{figure*}

Given the results of our tests on likelihoods using end-to-end simulations, showing potentially moderate affects of bias, we performed further investigations on the likelihoods as detailed here.  The aim was to distinguish between (i) issues arising from the noise in the simulations being different/more complicated than that described by the NCMs and (ii) issues in the likelihood schemes themselves.  Therefore we generated constraints from new sets of noise realisations, with the noise this time being generated as Gaussian realisations directly from the NCMs themselves. The new realisations are co-added with the same Gaussian realisations of the CMB generated for the tests performed in Sec.\ \ref{sec:results_sims}. 

The results for the two-dimensional tests are shown in Fig.\ \ref{fig:simulation_test_Gaussian}, for \momento to the left and \pixlike to the right. Each plot shows 100 posteriors for $r=0$ and $r=0.3$ on Gaussian realisations of the respective NCMs. We use two sets of NCMs: (i) in Figs. \ref{fig:simulation_test_momento_Gaussian_full} and \ref{fig:simulation_test_pixlike_Gaussian_full} the NCM for full-mission maps, called $100\times 143$ and (ii) in Figs. \ref{fig:simulation_test_momento_Gaussian_HM} and \ref{fig:simulation_test_pixlike_Gaussian_HM} the half-mission NCM obtained from the foreground cleaning procedure, labelled HM1, HM2 and HM1$\times$HM2. The former choice aligns with NCMs used in \citetalias{Belsunce21} while the  latter one corresponds with the work done in \citetalias{Belsunce22} and this paper, including the propagation of foreground-cleaning uncertainties. 

In both situations the likelihoods agree qualitatively and perform sensibly, not showing any strong bias. For posteriors in the top row, based on  $100\times 143$ spectra, for $r=0$ we obtain 43 posteriors for \momento and 45 and 47 posteriors for \pixlike out of the 100 that peak above zero. In the bottom row, with a significant tensor contribution of $r=0.3$, we obtain the expected behaviour, as previously observed on the end-to-end simulations, that the posteriors move to higher values of $r$. We get 61 posteriors for \momento and 47 and 51 posteriors for \pixlike out of the 100 that peak above $r$ of $0.3$. The posteriors in the bottom row of Fig.~\ref{fig:simulation_test_Gaussian} show Gaussian realisations of the half-mission covariance matrices, which is the situation closest to the constraints on $r$ presented in the body of the paper. For this test we obtain the following posterior peak distributions: for \momento we obtain 43 posteriors peaking above zero and 52 above $r=0.3$ for input values of $r=0$ and $r=0.3$ respectively. For \pixlike we obtain 48 and 42 posteriors that peak above zero for HM1 when $r=0$ and 45 and 49 above $r=0.3$ for an input value of $r$ of 0.3 for HM2.

For both sets of NCMs and for both likelihoods, when there is no tensor signal it can be seen indeed that no posteriors strongly disfavour $r=0$.  We also note that the spread in the recovered posteriors on the Gaussian realisations of the half-mission NCMs, which involve a different data split and have the foreground cleaning uncertainties included, is larger than that for the full-mission noise maps. 

These tests on correlated simulations give us confidence in the methods and also show that one should not expect to be able to constrain $r$ from low-$\ell$ \Planck data to better than about $\sim 0.2$. 

\bsp	
\label{lastpage}
\end{document}